\documentclass[conference]{IEEEtran}
\pdfoutput=1
\IEEEoverridecommandlockouts

\usepackage{inconsolata}

\usepackage{cite}
\usepackage{amsmath,amssymb,amsfonts}
\usepackage{algorithmic}
\usepackage{graphicx}
\usepackage{textcomp}
\usepackage{xcolor}
\usepackage{booktabs}
\usepackage{subfig}
\usepackage{packages}
\usepackage[font=footnotesize]{caption}
\usepackage{balance}

\def\BibTeX{{\rm B\kern-.05em{\sc i\kern-.025em b}\kern-.08em
    T\kern-.1667em\lower.7ex\hbox{E}\kern-.125emX}}

\begin{document}

\pagestyle{plain}

\title{Galois Field Arithmetics for Linear Network Coding using AVX\,512 Instruction Set Extensions}

\author{\IEEEauthorblockN{
Stephan M.\ Günther\IEEEauthorrefmark{1}, Nicolas Appel\IEEEauthorrefmark{2}, Georg Carle\IEEEauthorrefmark{1}\\
\IEEEauthorrefmark{1}Chair of Network Architectures and Services, Department of Computer Science\\
\IEEEauthorrefmark{2}Chair of Astronautics, Department of Mechanical Engineering\\
Technische Universität München\\
Email: \{guenther, n.appel, carle\}@tum.de
}}

\maketitle

\begin{abstract}
Linear network coding~\cite{AhlswedeCaiLiYeung2000} requires arithmetic operations over Galois fields, more specifically over finite extension fields.
While coding over GF(2) reduces to simple XOR operations, this field is less preferred for practical applications of random linear network coding due to high chances of linear dependencies~\cite{Cooper2000} and therefore redundant coded packets.
Coding over larger fields such as GF(16) and GF(256) does not have that issue, but is significantly slower.
SIMD vector extensions of processors such as AVX\,2 on x86-based systems or NEON on ARM-based devices offer the potential to increase performance by orders of magnitude~\cite{PlankGreenanMiller2013,GuRiUt2014}.

In this paper we present an implementation of different algorithms and Galois fields based on the AVX\,512 instruction set extension and integrate it into the finite field library \emph{libmoepgf}~\cite{GuRiUt2014}.
We compare the performance of the new implementation to the reference implementation based on AVX\,2, showing a significant increase in throughput.
In addition, we provide a survey of the best possible coding performance offered by a variety of different platforms.
The updated finite field library is available under the LGPL at~\cite{libmoepgfavx512}.
\end{abstract}


\section{Introduction}
For linear network coding, a transmitter generates a linear combination of packets that were previously been received.
Generally speaking, to generate a single encoded packet a linear combination of $N$ source packets is required, which can be expressed as a matrix-vector multiplication over a given extension field.
The number of packets $N$ is generally referred to as the \emph{generation size}.
The arithmetic complexity to generate a single encoded packet is therefore in $\mathcal O (M\cdot N)$, where $M$ denotes the packet length in number of data words.
In case of \emph{random} linear network coding, the coefficients forming the coding vector are randomly drawn from a finite extensions field.
For those applications, the generation size $N$ is severely limited to a small number of packets, commonly less than 256, as larger generations imply higher worst-case delays in case decoding is not possible due to packet loss.
Therefore, the arithmetic complexity of individual operations dominates the computation time in those cases \cite{PedersenHeideVingelmannFitzek2013,WangLi2006,ParamanathanPedersenLucaniFitzekKatz2013}.
While naive approaches using general purpose registers or simple lookup tables are severely limited in throughput, performance can be increased by multiple orders of magnitude when using vector instructions \cite{PlankGreenanMiller2013,GuRiUt2014}.

In this paper we extend the finite field library \emph{libmoepgf}~\cite{GuRiUt2014} by implementations for the AVX\,512 instruction set extensions as offered by Skylake-X based Xeon processors as well as the upcoming Intel Ice Lake microarchitecture.
The latter introduces AVX\,512 for the first time to the mobile and low-power segment as well as to desktop computers.

We start in Section~\ref{Ohfo9lei} with a formal description of network coding operations.
The support for AVX\,512 of different microarchitectures is discussed in Section~\ref{FeiM9du4}.
Section~\ref{ate0kuWe} presents the new AVX\,512 kernel using compiler instrinsics and discusses necessary changes compared to the AVX\,2-based implementation.
In Section~\ref{oqu4She2}, we demonstrate the performance advantages of the new implementation for different finite fields and packet sizes, and present a survey of multiple different computing platforms and the best achievable results on each.
Section~\ref{OhR4cho8} concludes the paper.

\section{Arithmetics of linear network coding}
\label{Ohfo9lei}
A data packet of length $l\,\si{\bit}$ can be considered as a sequence of $M=\lceil l/n\rceil$ data words, where $n$ denotes the word size in \si{\bit}.
The individual words $a_i$ for $i\in\{1,2,\ldots,M\}$ are elements of the finite extension field $F_q$ with $q=2^n$ elements.
Of particular interest for (random) linear network coding are the finite extension fields with 2, 4, 16, and 256 elements which we refer to as GF(2), GF(4), GF(16), and GF(256).
Assuming that $l$ is a multiple of \SI{8}{\bit}, which is reasonable as memory of most computers is accessible byte-wise, $l/n$ is also guaranteed to be an integral value for those fields.
Note that for GF(2) words represent individual bits while for GF(16) a word is a \emph{nibble}, i.\,e., a data word of \SI{4}{\bit}.

A packet can therefore be written as vector $\mathbit{a} = [a_1,a_2,\ldots,a_M]^T$.
Given a generation of $N$ source packets, those packets form a matrix
\begin{align}
	\mathbit A =
	\left [
	\begin{matrix}
		\mathbit a_{1} & \ldots & \mathbit a_{N}
	\end{matrix}
	\right ]
	=
	\left [
	\begin{matrix}
		a_{11} & \ldots & a_{1N}\\
		\vdots & \ddots & \vdots\\
		a_{M1} & \ldots & a_{MN}
	\end{matrix}
	\right ] \in F_q^{M\times N}.
\end{align}
To generate an encoded packet $\mathbit b$, we need a coding vector $\mathbit c \in F_q^N$.
Depending on whether or not we talk about \emph{random} linear network coding, the components of $\mathbit c$ may be chosen identically and independently distributed from $F_q$ or by means of some deterministic algorithm.
The actual encoding operation now consists in the matrix-vector multiplication
\begin{align}
	\mathbit b = \mathbit A \mathbit c
	=
	\sum_{i=1}^N c_i \mathbit a_i. \label{Shog4ire}
\end{align}
Given the extension fields mentioned above, addition over $F_q$ is always a bit-wise XOR operation while multiplication is a modulo operation given a reduction polynomial specific to the respective field.

According to \eqref{Shog4ire}, we now need efficient algorithms to multiply a vector $\mathbit a$ by some constant value $c$ and to add the result to an array used as accumulator over the respective extension field.
That operation is commonly known as \emph{multiply and add (madd)}.
Such an algorithms suitable vector instruction set extensions has been proposed by Anvin~\cite{Anvin2004} and later by Plank et al.\ in~\cite{PlankGreenanMiller2013} and is referred to as \emph{shuffle} algorithm.
As its name implies, this algorithm requires a shuffle instruction that swaps words in vector registers.
Another algorithm called \emph{imul} has been proposed by us in~\cite{GuRiUt2014}.
It does not require any special instructions, but its complexity linearly depends on the word size, i.\,e., the larger the word size the more operations are required.
Both algorithms are implemented for SSSE\,3, AVX\,2 and NEON instruction set extensions in libmoepgf~\cite{GuRiUt2014}.

We now continue to extend those implementations to the AVX\,512 instruction set extensions, which double the register width and\,---\,in theory\,---\,provide a two times speedup compared to the AVX\,2 implementations.

\section{AVX\,512 support}
\label{FeiM9du4}
AVX\,512 is a family of instruction set extensions widely introduced by Intel with the Skylake-X server processors in 2017.
Beforehand, a subset of AVX\,512 extensions was available in Intel's Xeon Phi processors, which were derived from the Larrabee project \cite{Seiler:2008:LMX:1360612.1360617}.
Skylake-X processors thereby vary in the amount of execution units for fused multiply-and-add (FMA).
In addition, Intel uses rather complex mechanisms to determine the maximum operating frequency depending on the number of active cores, the instruction set extensions in use, and even specific instructions being executed.

\subsection{Variants of AVX\,512}\label{fuo0Ahb4}
As mentioned before, AVX\,512 is not a single instruction set extension but a whole family of extensions.
The relevant parts for the scope of this paper are \emph{AVX\,512-F (foundation)}, which is supported by all processors supporting AVX\,512, as well as \emph{AVX\,512-BW (byte and word)}, which is so far only supported by Skylake-X and Ice Lake processors.
Without AVX\,512-BW, there is no support for byte-wise shuffle operations which makes a port of the AVX\,2 shuffle algorithm impossible.
However, the imul algorithm would still be useable.

Since AVX\,512-BW is supported by all Skylake-X and Ice Lake processors, this is a rather theoretic case suitable only for the discontinued Xeon Phi processors and certain add-in accelerators.
Nevertheless, libmoepgf differentiates between these feature sets and enables an AVX\,512-F variant of the imul algorithm if necessary.

\begin{table}
	\centering
	\vspace{1.2ex}
	\caption{Maximum single-core turbo frequencies in \si{\giga\hertz} of different processors when executing normal, AVX\,2, AVX\,512, and NEON code}
	\begin{tabular}{lrrrrr}
		\toprule
		\textbf{CPU}     & \textbf{normal}       & \textbf{AVX\,2}       & \textbf{AVX\,512} & \textbf{NEON}\\
		\midrule
		Intel Xeon Gold 6130   & 3.7 & \footnotemark[1]{}3.7\,/\,3.6 & \footnotemark[1]{}3.6\,/\,3.5 & --\\
		Intel Xeon Silver 4116 & 3.0 & \footnotemark[1]{}3.0\,/\,2.9 & \footnotemark[1]{}2.9\,/\,1.8 & --\\
		Intel Xeon D-2166\,NT  & 3.0 & \footnotemark[1]{}3.0\,/\,2.9   & \footnotemark[1]{}2.9\,/\,2.8   & --\\
		Intel Xeon E5-2696\,v3 & 3.8 & 3.4 & --                  & -- \\
		Intel Xeon E5-2650     & 2.8 & -- & -- & -- \\
		AMD Epyc 7601          & 3.2 & \footnotemark[2]{}3.2 & -- & --\\
		AMD Ryzen 1700         & 3.7 & \footnotemark[2]{}3.5 & -- & --\\
		Broadcom BCM\,2711       & 1.5 & --                    & --                    & 1.5\\
		\bottomrule
	\end{tabular}
	\label{Iecho0zi}
\end{table}

\footnotetext[1]{Maximum turbo depending on whether light/heavy instructions are being executed.}
\footnotetext[2]{Register width is limited to \SI{128}{\bit}, i.\,e., \SI{256}{\bit} operations are split into two operations.}

\subsection{Special frequency ranges for AVX code}
\label{oozaiQu7}
Intel has been using different maximum frequencies for their processors since AVX\,2.
Unfortunately, Intel does not make the clocking behaviour of their processors transparent to end users and publish those details only as part of their specification update manual \cite{intel_xeon-sepc_update}.
Assuming that the CPU is not thermally limited or otherwise constrained, current Skylake-X based processors have three different frequency levels referred to as \emph{license levels (LVL)} \cite{intel_skylake-x}.
Within a given LVL, the number of active cores further restricts the maximum frequency.
We do not further consider multiple active cores in this paper since we limit our benchmarks to a single thread.
LVL\,0 refers to the fastest turbo frequencies applicable to normal code, LVL\,1 is slower and applies to most AVX\,2 instructions, while LVL\,2 is the slowest level.

Whether an AVX\,512 instruction is executed within the envelope of LVL\,1 or LVL\,2 depends on whether it is a so called \emph{light} or \emph{heavy} instruction.
The latter group are those instructions being executed on the FMA units while the former are most logic and byte-wise instructions, in particular the shuffle operation.

Depending on the processor model, the maximum frequencies between LVL\,1 and LVL\,2 might differ significantly.
Since the imul algorithm is based on packed integer multiplications, which are considered heavy instructions, the processor runs much slower compared to the shuffle algorithm that only uses light instructions.
Table~\ref{Iecho0zi} shows these frequencies for the processors we use for comparison in this paper.
While the Intel Xeon Gold 6130 reduces its frequency by just \SI{200}{\mega\hertz} when executing AVX\,512 heavy code on a single core compared to normal code, the Intel Xeon Silver 4116 throttles by \SI{1200}{\mega\hertz} under the same conditions \cite{intel_xeon-sepc_update}.
Consequently, the gain of AVX\,512 enabled code may be much less than one might expect.

\vspace{\baselineskip}
AMD has so far no special operating frequencies when executing AVX code but the  AMD Epyc 7xx1 series only supports \SI{128}{\bit} operations, i.\,e., AVX\,2 operations are split into two \SI{128}{\bit} operations.
This changed with the 7xx2 series, but there is still no support AVX\,512.

However, we found that our AMD Ryzen 1700, which has a maximum single-core turbo of \SI{3.7}{\giga\hertz}, reduces its clock speed to \SI{3.5}{\giga\hertz} under heavy load, i.\,e., it did not achieve its maximum advertised single-core turbo independent of the load offered while it was neither limited by thermal nor power constraints.

\section{Implementation}
\label{ate0kuWe}
We implement AVX\,512 variants of both the shuffle and imul algorithms for GF(2), GF(4), GF(16), and GF(256).
The following sections show the exemplary implementations for GF(4) as they contain all special cases of the larger fields but allow for a more compact representation since less constant values are required.
The algorithms are integrated into libmoepgf and its benchmark application.

\subsection{shuffle algorithm}
Algorithm~\ref{alg:shuffle} shows the implementation of the shuffle algorithm using C compiler instrinsics to make sure proper AVX\,512 code is produced.
The function expects two regions \texttt{a} and \texttt{b} corresponding to a source packet $\mathbit{a}_i$ and the accumulator array $\mathbit b$ as well as a coding coefficient $c_i$ represented by the constant value \texttt{c}.
It then calculates $\mathbit b := \mathbit b + c_i\cdot\mathbit a_i$.
The register variables in Algorithm~\ref{alg:shuffle} are preloaded with various constants, in particular the lookup tables and bit masks needed for the shuffle algorithm.
Afterwards, the two trivial cases for $c_i \in \{0,1\}$ are caught, which result in either no operation or a simple XOR between the input arrays.
Otherwise, temporary registers are preloaded and the shuffle algorithm performs the respective madd operation.

\lstset{
    language=c,
    basicstyle=\scriptsize\ttfamily,
    framerule=.5pt,
    frame=lines,
    belowcaptionskip=3pt,
    abovecaptionskip=-3pt,
    keywordstyle=\color{black}\bfseries,
    morekeywords={uint8_t,uint64_t,int,size_t}
}
\renewcommand\lstlistingname{Algorithm}
\begin{lstlisting}[caption={shuffle for GF(4) using AVX\,512-BW},label={alg:shuffle}]
void
madd4_avx512bw(uint8_t *b, uint8_t *a, uint8_t c, size_t len)
{
  uint8_t *end;
  register __m512i in1, in2, out, t1, t2, m1, m2, l, h;
  register __m128i bc;

  if (c == 0)
    return;

  if (c == 1) {
    xorr_avx512(b, a, length);
    return;
  }

  bc = _mm_load_si128((void *)tl[c]);
  t1 = _mm512_broadcast_i32x4 (bc);
  bc = _mm_load_si128((void *)th[c]);
  t2 = _mm512_broadcast_i32x4 (bc);
  m1 = _mm512_set1_epi8(0x0f);
  m2 = _mm512_set1_epi8(0xf0);

  for (end=b+length; b<end; b+=64, a+=64) {
    in2 = _mm512_load_si512((void *)a);
    in1 = _mm512_load_si512((void *)b);
    l = _mm512_and_si512(in2, m1);
    l = _mm512_shuffle_epi8(t1, l);
    h = _mm512_and_si512(in2, m2);
    h = _mm512_srli_epi64(h, 4);
    h = _mm512_shuffle_epi8(t2, h);
    out = _mm512_xor_si512(h,l);
    out = _mm512_xor_si512(out, in1);
    _mm512_store_si512((void *)b, out);
  }
}
\end{lstlisting}

Similar implementations exist for the other extension fields.
The main difference consists in differing values for the lookup tables and bit masks.
Note that this implementation relies on instruction available only in the AVX\,512-BW subset of AVX\,512 as \texttt{\_mm512\_shuffle\_epi8} belongs to the byte-wise instructions.
For microarchitectures not implementing AVX\,512-BW, this algorithm is not applicable.

\subsection{imul algorithm}
On platforms \emph{not} supporting AVX\,512-BW but AVX\,512-F (foundation), the imul algorithm shown in Algorithm~\ref{alg:imul} is used instead.
The algorithm again catches trivial cases first and then sets up a number of temporary registers with constant values:
\texttt{mi} contains bit masks used to isolate individual coefficients within a word and are repeated as many times as words fit into the \SI{512}{\bit} registers.
\texttt{sp} contains the powers of the constant factor $c_i$ for the respective extension field.
Afterwards, the loop performs literally the required polynomial multiplications on $c_i$ and $\mathbit b_i$ using only AVX\,512-F instructions.
The result is XORed into the destination register, which corresponds to the add-part of the madd operation.

\lstset{
    language=c,
    basicstyle=\scriptsize\ttfamily,
    framerule=.5pt,
    frame=lines,
    belowcaptionskip=3pt,
    abovecaptionskip=-3pt,
    keywordstyle=\color{black}\bfseries,
    morekeywords={uint8_t,uint64_t,int,size_t}
}
\renewcommand\lstlistingname{Algorithm}
\begin{lstlisting}[caption={imul for GF(4) using AVX\,512-F},label={alg:imul}]
void
madd4_avx512f(uint8_t *b, uint8_t *a, uint8_t c, size_t len)
{
  uint8_t *end;
  register __m512i reg1, reg2, ri[2], sp[2], mi[2];
  const uint8_t *p = pow[c];

  if (c == 0)
    return;

  if (c == 1) {
    xorr_avx512(b, a, length);
    return;
  }

  mi[0] = _mm512_set1_epi8(0x55);
  mi[1] = _mm512_set1_epi8(0xaa);
  sp[0] = _mm512_set1_epi32(p[0]);
  sp[1] = _mm512_set1_epi32(p[1]);

  for (end=b+length; b<end; b+=64, a+=64) {
    reg2 = _mm512_load_si512((void *)a);
    reg1 = _mm512_load_si512((void *)b);
    ri[0] = _mm512_and_si512(reg2, mi[0]);
    ri[1] = _mm512_and_si512(reg2, mi[1]);
    ri[1] = _mm512_srli_epi32(ri[1], 1);
    ri[0] = _mm512_mullo_epi32(ri[0], sp[0]);
    ri[1] = _mm512_mullo_epi32(ri[1], sp[1]);
    ri[0] = _mm512_xor_si512(ri[0], ri[1]);
    ri[0] = _mm512_xor_si512(ri[0], reg1);
    _mm512_store_si512((void *)b, ri[0]);
  }
}
\end{lstlisting}

While this algorithm is guaranteed to work on all devices supporting AVX\,512, it has two drawbacks.
First, its complexity depends on the word length as the number of multiplications and logical operations within the main loop depend on the word length of the respective extension field.
Second, the integer multiplications used in the main loop are heavy instructions and therefore subject to lower frequencies as discussed in Section~\ref{oozaiQu7}.

\begin{figure*}[t]
	\centering
	\subfloat[GF(256)]{\label{kaeR8eew-1}
	\includegraphics[width=.49\textwidth]{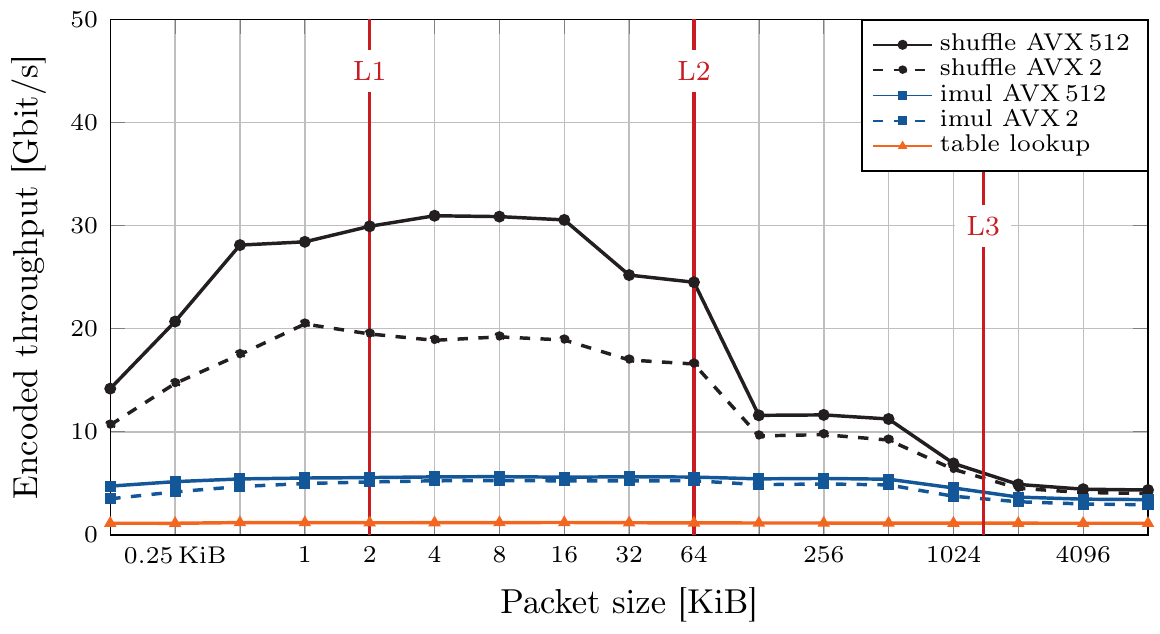}}
	\hfill%
	\subfloat[GF(16)]{\label{kaeR8eew-2}
	\includegraphics[width=.49\textwidth]{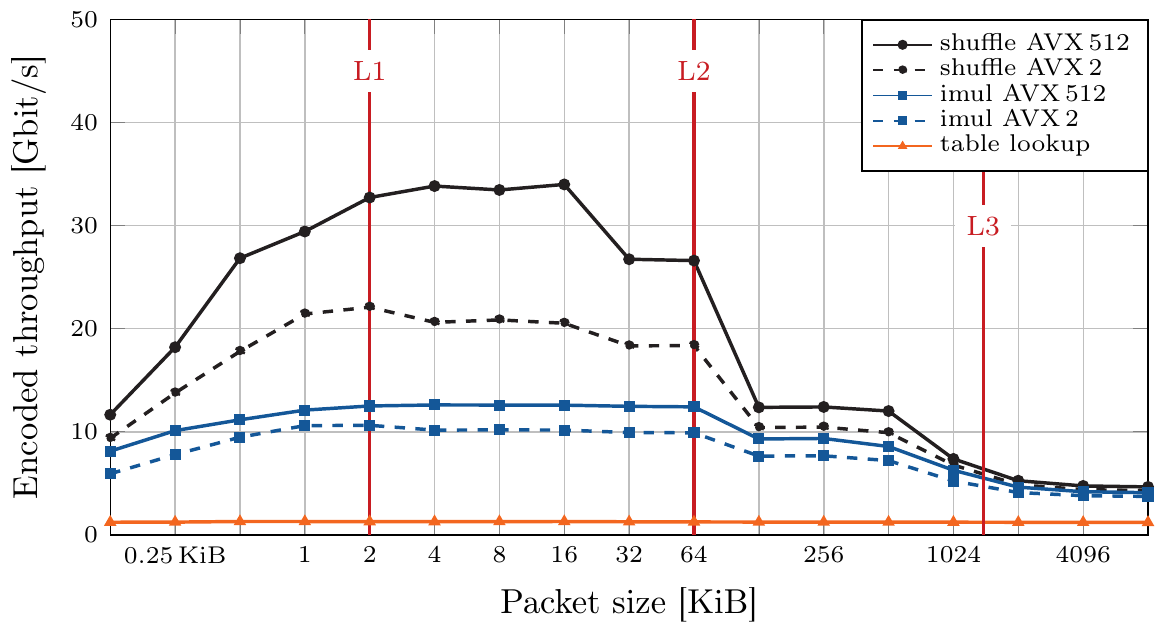}}

	\subfloat[GF(4)]{\label{kaeR8eew-3}
	\includegraphics[width=.49\textwidth]{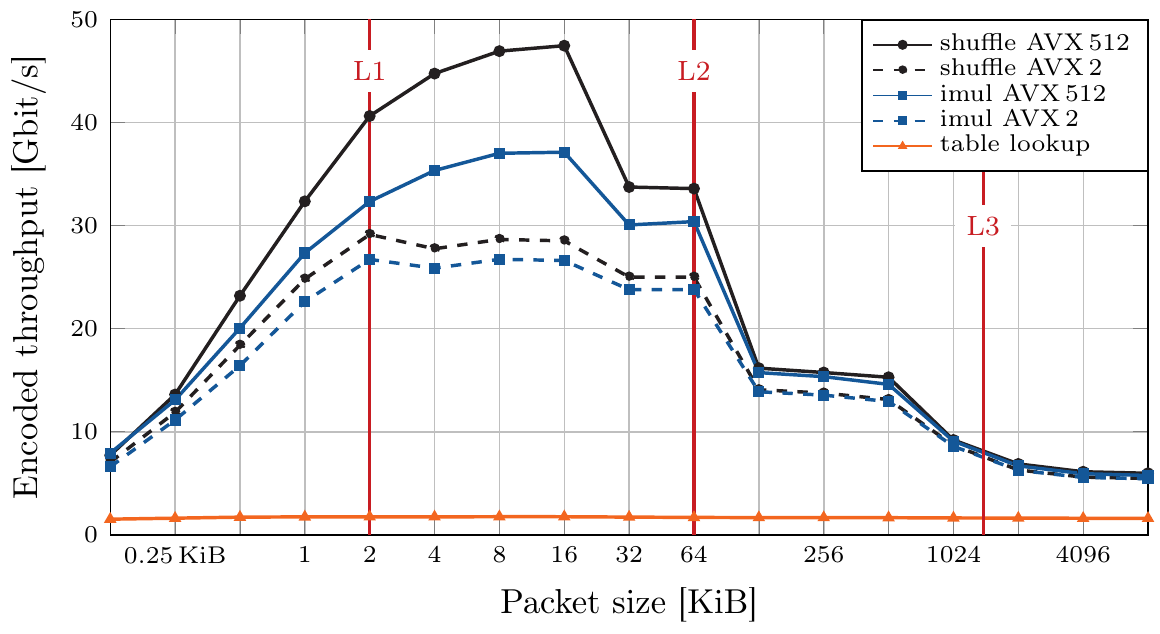}}
	\hfill%
	\subfloat[GF(2)]{\label{kaeR8eew-4}
	\includegraphics[width=.49\textwidth]{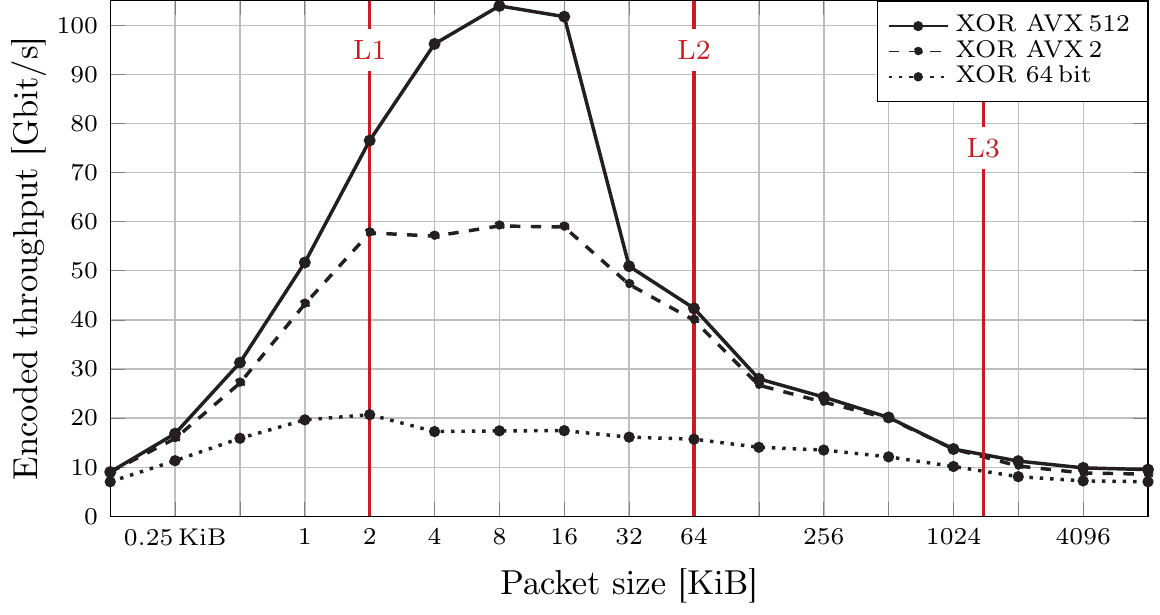}}
	\caption{Throughput in \si{\giga\bit\per\second} is the amount of random linear combinations of 16 packets at varying packet size from \SI{128}{\byte} up to \SI{8}{\kibi\byte} on an Intel Xeon Gold 6130 with one thread pinned to a specific core.}
	\label{kaeR8eew}
\end{figure*}

\section{Evaluation}
\label{oqu4She2}
We first evaluate the implementation of the shuffle and imul algorithms for AVX\,512 vs.\ the AVX\,2 variant proposed in~\cite{GuRiUt2014} on an Intel Xeon Gold 6130 CPU in Section~\ref{IeToo1uw}, showing significant performance advantages when making use of AVX-512BW in particular.
Section~\ref{Ahs4ooch} contains a survey of the best performing settings for different processors, showing the microarchitectural advances over the past few years.

\subsection{Performance depending on ISA extensions}\label{IeToo1uw}
Figure~\ref{kaeR8eew} shows the test results on an Intel Xeon Gold 6130.
The process is pinned to a specific core of the CPU. 
Each measurement point is averaged over $4\cdot 10^7$ repetitions to preclude any short-term effects from influencing the measurement.
Figures~\ref{kaeR8eew-1}~--~\ref{kaeR8eew-4} show the encoding throughput with varying packet sizes ranging from \SI{256}{\byte} up to \SI{8}{\kibi\byte} for the different Galois fields.
The encoding throughput is thereby defined as the amount of encoded packets measured in \si{\giga\bit\per\second} based on 16 uncoded source packets of the respective size that are randomly combined, i.\,e., the test resembles random linear network coding with a generation size of 16.
The vertical red lines indicate the packet sizes at which the working set fills the L1, L2, and L3 caches, respectively.
For instance, the Intel Xeon Gold 6130 has \SI{1}{\mebi\byte} of L2 cache per core which means that 16 packets of \SI{64}{\kibi\byte} exactly fill the L2 cache, which is indicated by the middle red line.\footnotemark

\footnotetext{%
To be precise, the working set consists of 16 uncoded source packets plus one array of packet size used as accumulator for the encoded output.
Furthermore, the L2 Cache of Skylake-X is inclusive, i.\,e., the contents of both the L1 data and instruction cache (\SI{32}{\kibi\byte} each) are always present in the L2 cache.
}

The results for GF(256) (Figure~\ref{kaeR8eew-1}) and GF(16) (Figure~\ref{kaeR8eew-2}) clearly show a significant drop in performance once the working set exceeds the L2 cache size.
While the results for GF(4) (Figure~\ref{kaeR8eew-3}) and in particular GF(2) (Figure~\ref{kaeR8eew-4}) also drop at the L2 cache boundary, there is another significant drop at a packet size of \SI{32}{\kibi\byte}.
A slight drop of performance at or (depending on the cache size) even before the cache boundary can be a result of the L2 cache being inclusive and the working set consisting not only of 16 source packet but also an array used as accumulator for the encoded packet, such significant drops must have other reasons.
Interestingly, this phenomenon was neither observed on the Haswell-based processor of our original publication \cite{GuRiUt2014}, nor is it reproducible with any processor other than those based on Skylake-X -- even the desktop derivatives of Skylake (not Skylake-X) do not show this behavior.
At the moment, we do not have a conclusive explanation for this behaviour, but we can preclude errors in the benchmark itself (since the other processors behave as expected) and also any mitigations of security issues on Intel processors (Spectre, Meltdown etc.) as we conducted the tests both with and without those mitigations showing virtually no difference in performance.

In any case, performance is severely limited by memory bandwidth resulting in a significant drop once the L2 cache is exceeded.
As we expect, most affected are fast algorithms using AVX\,2 and AVX\,512 demonstrating a massive processing power of the CPU as long as execution units are kept busy.
On the other hand, too small packet sizes result in too much overhead, e.\,g.\ start and end code for loops.
For the commonly used GF(256), maximum performance is achieved for packet sizes between \SI{512}{\byte} and \SI{16}{\kibi\byte}, which covers a significant range of common packet sizes for networking applications.

The reason why the encoding throughput over GF(4) is considerable higher compared to GF(16) and GF(256) is that in case of GF(4) half of the operations are trivial:
in case of multiplication by $0$ or $1$ (see Section~\ref{ate0kuWe}), the operations reduce to a null or copy/XOR operation, and as there are only four elements in GF(4), these cases make up half of all operations.

Although it is obvious, it should be noted that increasing the generation size of course increases the working set, e.\,g.\ doubling the generation size cuts the packet size at which the algorithms become bandwidth limited in half.

\vspace{\baselineskip}
Given that modern CPUs are able to encode at speeds of \SI{30}{\giga\bit\per\second} over GF(256) using a single core only, the encoding operations do not pose a significant overhead on those systems anymore.
However, on older processors or embedded systems, things might look different which is discussed in the following section.

\subsection{Comparison between CPUs}\label{Ahs4ooch}
To compare the performance of different CPUs, we evaluate our library on eight different processors listed in Table~\ref{Iecho0zi}.
Since the performance depends on the two major factors\,---\,SIMD support (and therefore the chosen algorithm)  and packet size (and therefore cache size)\,--\,we define the following benchmark procedure:
\begin{enumerate}
	\item Each CPU is benchmarked with the fastest algorithm available, i.\,e., making use of the specific SIMD extensions available.
	\item The result for each CPU is the average of results for working sets fitting in the L2 cache.
\end{enumerate}

\balance

\begin{figure}
	\centering
	\includegraphics[width=.5\textwidth]{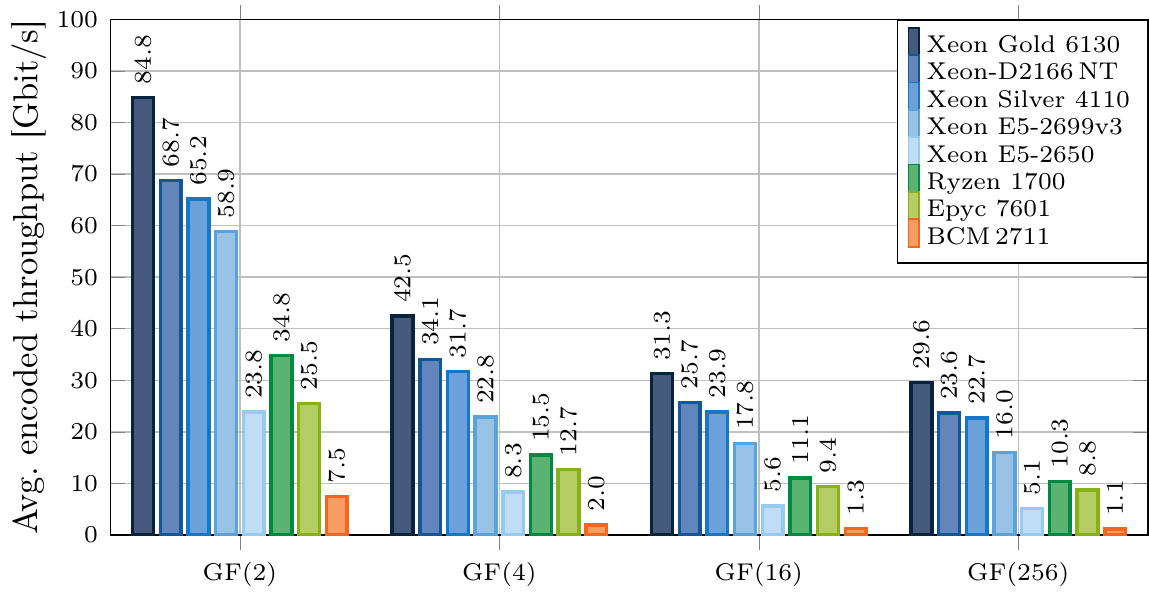}
	\caption{Encoding throughput measured in \si{\giga\bit\per\second} over different Galois fields using a variety of CPUs.
For each CPU we choose the fastest algorithm possible and average the encoding throughput over all packet sizes until the working set approaches the L2 capacity.}
	\label{Ees6zahs}
	\vspace{-2ex}
\end{figure}

The results are shown in Figure~\ref{Ees6zahs}.
The first three bars for each Galois field show the results of recent Intel Xeon processors, which are the only processors under test supporting the new AVX\,512 instruction set extensions.
According to Table~\ref{Iecho0zi}, the faster Xeon Gold 6130 uses a frequency of \SI{3.6}{\giga\hertz} compared to \SI{2.9}{\giga\hertz} of the Xeon-D2166 and Xeon Silver 4116 because the shuffle algorithm is preferred in all cases over the imul algorithm.
Consequently, the difference in encoding throughput is almost solely based on the difference in clock frequency.
The fact that the Xeon-D2166 is slightly faster than the Xeon Silver 4116 albeit identical clock speeds is most likely related to some background tasks on the latter because we were unable to dedicate the machine for benchmarks.

The fourth bar represents the aging Intel Xeon E5-2696\,v3, which is a processor based on the Haswell microarchitecture, i.\,e., two generations before Skylake.
Considering that this processor does not support AVX\,512 but only AVX\,2, the results are exceptionally good.

The fifth bar shows the results for a Xeon 2650, which is based on the Sandy Bridge EP microarchitecture from 2012 and thus a legacy system.
It does not even support AVX\,2 and therefore libmoepgf falls back to SSE\,2\,/\,SSSE\,3 instruction set extensions.

Column number six and seven represent an AMD Ryzen 1700 desktop and AMD Epyc 7601 server processor.
Under perfect test conditions one would expect that the difference is solely due to clock speed since both processors use the exactly same CPU core complexes.
However, the Epyc falls back significantly which we believe is caused by background load on this machine as we were not able to dedicate this machine for benchmark purposes.
Compared to the faster Intel CPUs, we have to keep in mind that these AMD processors only have \SI{128}{\bit} registers.
Consequently, AVX\,2 operations have to be split into two operations and AVX\,512 is not supported at all.
The newer Epic 7xx2 and Ryzen 3000 CPUs are expected to be on par with Intel CPUs supporting AVX\,2.

As a comparison for current embedded systems, we also include the performance on a Broadcom BCM\,2711 as used on the Raspberry Pi Model 4B, which is an ARMv8-based CPU supporting the NEON instruction set extensions.
Considering that this CPU has not even a passive cooler and is clocking at only \SI{1.5}{\giga\hertz}, the results are quite astonishing: even over GF(256) we achieve over \SI{1}{\giga\bit\per\second} of encoded throughput.

\clearpage
\section{Conclusion}
\label{OhR4cho8}
We can conclude from the results that increases in encoding throughput almost solely stem from advances in SIMD support and clock frequency.
Main memory throughput is of less importance due to the ratio between packet and cache sizes.
In case of memory limitations, AVX\,2 or even AVX\,512 support does not offer a significant benefit as shown in Section~\ref{fuo0Ahb4}.

However, within the range of packet sizes interesting for networking application we see a substantial increase in encoding throughput when using AVX\,512 extensions.
The increase is not a perfect two-time scaling compared to AVX\,2, which is partially but not completely explained by lower frequencies when AVX\,512 code is executed.
A similar phenomenon was already observed in~\cite{GuRiUt2014} when comparing SSSE\,3 to AVX\,2 implementations.

The new implementations are integrated into libmoepgf, which is published under the LGPL at \cite{libmoepgfavx512}.
Of course we are open for suggestions regarding the improvement of libmoepgf.

\bibliographystyle{IEEEtran}
{
\footnotesize
\bibliography{IEEEabrv,lit.bib}

\begin{thebibliography}{10}
\providecommand{\url}[1]{#1}
\csname url@samestyle\endcsname
\providecommand{\newblock}{\relax}
\providecommand{\bibinfo}[2]{#2}
\providecommand{\BIBentrySTDinterwordspacing}{\spaceskip=0pt\relax}
\providecommand{\BIBentryALTinterwordstretchfactor}{4}
\providecommand{\BIBentryALTinterwordspacing}{\spaceskip=\fontdimen2\font plus
\BIBentryALTinterwordstretchfactor\fontdimen3\font minus
  \fontdimen4\font\relax}
\providecommand{\BIBforeignlanguage}[2]{{%
\expandafter\ifx\csname l@#1\endcsname\relax
\typeout{** WARNING: IEEEtran.bst: No hyphenation pattern has been}%
\typeout{** loaded for the language `#1'. Using the pattern for}%
\typeout{** the default language instead.}%
\else
\language=\csname l@#1\endcsname
\fi
#2}}
\providecommand{\BIBdecl}{\relax}
\BIBdecl

\bibitem{AhlswedeCaiLiYeung2000}
R.~Ahlswede, N.~Cai, S.-Y. Li, and R.~Yeung, ``{N}etwork {I}nformation
  {F}low,'' \emph{IEEE Transactions on Information Theory}, vol.~46, no.~4, pp.
  1204--1216, Jul. 2000.

\bibitem{Cooper2000}
C.~Cooper, ``{O}n the {D}istribution of {R}ank of a {R}andom {M}atrix over a
  {F}inite {F}ield,'' \emph{Random Struct. Algorithms}, vol.~17, no. 3-4, pp.
  197--212, Oct. 2000.

\bibitem{PlankGreenanMiller2013}
J.~Plank, K.~Greenan, and E.~Miller, ``{S}creaming {F}ast {G}alois {F}ield
  {A}rithmetic using {I}ntel {SIMD} {I}nstructions,'' \emph{USENIX Conference
  on File and Storage Technologies}, vol.~11, Aug. 2013.

\bibitem{GuRiUt2014}
S.~M. G\"{u}nther, M.~J. Riemensberger, and W.~Utschick, ``{E}fficient
  {\sc{gf}} {A}rithmetic for {L}inear {N}etwork {C}oding using {H}ardware
  {\sc{simd}} {E}xtensions,'' in \emph{IEEE International Symposium on Network
  Coding (NetCod)}, Aalborg, Denmark, 7 2014.

\bibitem{libmoepgfavx512}
S.~M. G{\"u}nther and N.~Appel, ``Supplemental material: libmoepgf,''
  https://moep80211.net/plink/libmoepgf-avx512.

\bibitem{PedersenHeideVingelmannFitzek2013}
M.~Pedersen, J.~Heide, P.~Vingelmann, and F.~Fitzek, ``Network {C}oding over
  the $2^{32}-5$ {P}rime {F}ield,'' in \emph{IEEE International Conference on
  Communications}, Jun. 2013, pp. 2922--2927.

\bibitem{WangLi2006}
M.~Wang and B.~Li, ``{H}ow {P}ractical is {N}etwork {C}oding?'' in \emph{IEEE
  International Workshop on Quality of Service}, Jun. 2006, pp. 274--278.

\bibitem{ParamanathanPedersenLucaniFitzekKatz2013}
A.~Paramanathan, M.~Pedersen, D.~Lucani, F.~Fitzek, and M.~Katz, ``{L}ean and
  {M}ean: {N}etwork {C}oding for {C}ommercial {D}evices,'' \emph{IEEE Wireless
  Communications}, vol.~20, no.~5, pp. 54--61, Oct. 2013.

\bibitem{Anvin2004}
H.~Anvin, ``{T}he {M}athematics of {RAID}-6,'' 2004.

\bibitem{Seiler:2008:LMX:1360612.1360617}
\BIBentryALTinterwordspacing
L.~Seiler, D.~Carmean, E.~Sprangle, T.~Forsyth, M.~Abrash, P.~Dubey,
  S.~Junkins, A.~Lake, J.~Sugerman, R.~Cavin, R.~Espasa, E.~Grochowski,
  T.~Juan, and P.~Hanrahan, ``Larrabee: A many-core x86 architecture for visual
  computing,'' \emph{ACM Trans. Graph.}, vol.~27, no.~3, pp. 18:1--18:15, Aug.
  2008. [Online]. Available: \url{http://doi.acm.org/10.1145/1360612.1360617}
\BIBentrySTDinterwordspacing

\bibitem{intel_xeon-sepc_update}
\emph{{I}ntel {X}eon {P}rocessor {S}calable {F}amily -- {S}pecification
  {U}pdate}, Intel, Jun. 2019.

\bibitem{intel_skylake-x}
\emph{{I}ntel 64 and {IA}-32 {A}rchitectures {O}ptimization {R}eference
  {M}anual}, Intel, Apr. 2019.

\end{thebibliography}
}

\end{document}